# Universality of Coherent Raman Gain Suppression in Gas-Filled Broadband-Guiding Photonic Crystal Fibers

P. Hosseini[1], M. K. Mridha[1], D. Novoa[1], A. Abdolvand[1], and P. St.J. Russell[1,2]

[1]*Max Planck Institute for the Science of Light, Staudt-Str. 2, 91058 Erlangen, Germany*
[2]*Department of Physics, University of Erlangen-Nuremberg, Germany*

As shown in the early 1960s, the gain in stimulated Raman scattering (SRS) is drastically suppressed when the rate of creation of phonons (via pump-to-Stokes conversion) is exactly balanced by the rate of phonon annihilation (via pump-to-anti-Stokes conversion). This occurs when the phonon coherence waves—synchronized vibrations of a large population of molecules—have identical propagation constants for both processes, i.e., they are phase-velocity matched. As recently demonstrated, hydrogen-filled photonic crystal fiber pumped in the vicinity of its zero-dispersion wavelength provides an ideal system for observing this effect. Here we report that Raman gain suppression is actually a universal feature of SRS in gas-filled hollow-core fibers, and that it can strongly impair SRS even when the dephasing rate is high, particularly at high pump powers when it is normally assumed that nonlinear processes become more (not less) efficient. This counter-intuitive result means that inter-modal stimulated Raman scattering (for example between $LP_{01}$ and $LP_{11}$ core modes) begins to dominate at high power levels. The results reported have important implications for fiber-based Raman shifters, amplifiers or frequency combs, especially for operation in the ultraviolet, where the Raman gain is much higher.

## I. INTRODUCTION

Stimulated Raman scattering (SRS) has been used for efficient generation of new optical frequencies in several different configurations, including gas cells [1], solid-state materials [2], silicon waveguides [3] and glass-core fibers [4,5]. Many practical applications have emerged from glass-fiber-based SRS, where tight field confinement and long interaction lengths permit much lower threshold powers. In addition, the ability to tailor the dispersion has allowed reliable observation of effects such as the soliton self-frequency shift [6]. Hollow-core photonic crystal fiber (HC-PCF) offers similar advantages in the context of gas-based SRS, and so is an ideal vehicle for studying all kinds of gas-laser interactions [7,8]. One particular type of HC-PCF guides by anti-resonant reflection (ARR), offering low loss transmission over a very broad wavelength range, as well as pressure-tunable dispersion when filled with gas. ARR-PCFs have enabled the generation of bright tunable vacuum and deep ultraviolet (UV) light [9], Raman frequency combs [10,11], and supercontinua extending from the vacuum UV to the infrared [12]. In addition, the inherent multi-mode nature of most ARR-PCFs [13, 14] has been exploited to up-convert broadband signals [15] and to enhance the generation of anti-Stokes sidebands in the UV [16].

Being seeded from noise, pump-to-Stokes conversion is automatically phase-matched, giving rise to a phonon coherence wave with wavevector $(\beta_P - \beta_S)$ where $\beta_P$ and $\beta_S$ are the propagation constants of the pump and the Stokes waves respectively. Pump-to-anti-Stokes conversion, on the other hand, requires a coherence wave with wavevector $(\beta_{AS} - \beta_P)$, where AS denotes anti-Stokes. It might be expected, therefore, that anti-Stokes generation will be strongest when the two coherence waves are identical, i.e., $\vartheta = (\beta_S + \beta_{AS}) - 2\beta_P = 0$. However, as first predicted by Bloembergen and Shen [17] and subsequently studied by several groups [18-22], the Raman gain is actually dramatically suppressed when $\vartheta = 0$. This is because the rate of creation of phonons (via pump-to-Stokes conversion) is exactly balanced by the rate of phonon annihilation (via pump-to-anti-Stokes conversion), resulting in suppression of the Raman gain.

This effect was recently observed in $H_2$-filled kagomé-type ARR-PCF, pressure-tuned in the vicinity of its zero dispersion point (ZDP) so that $\vartheta = 0$ for intramodal SRS between $LP_{01}$-like core modes [23]. Although in this way intra-$LP_{01}$-mode Raman gain was strongly suppressed, intermodal SRS between $LP_{01}$-like pump and $LP_{02}$-like Stokes turned out to be strongly enhanced, through absence of gain suppression.

Here, we report that gain suppression is in fact a universal feature of SRS in broadband-guiding HC-PCFs, and that, for high enough pump power, it can be observed even when the system is operated far from $\vartheta = 0$. We find that the Raman gain can actually decrease with increasing pump power, because nonlinear coupling between Stokes, pump and anti-Stokes signals overcomes the dephasing effect of $|\vartheta| > 0$. This has important practical implications for the design of high-power Raman shifters, amplifiers and frequency-comb sources based on hollow-core fiber technology. The effect is even more pronounced in the ultraviolet spectral region, where the Raman gain is much higher.

## II. ANALYSIS OF SYSTEM

The propagation constant of the $LP_{ij}$-like core mode at frequency $\omega$ can be expressed to a good approximation in the form [24]:

$$\beta^{ij} = \sqrt{k_0^2 n_g^2(p,\omega,T) - u_{ij}^2/a^2(\omega)} \qquad (1)$$

where $k_0 = \omega/c$ is the vacuum wavevector, $c$ is the speed of light in vacuum, $n_g(p,\omega,T) = (1+\delta(\omega)pT_0/p_0T)^{0.5}$ is the refractive index of the filling gas at pressure $p$ and temperature $T$ and $\delta(\omega)$ is the Sellmeier expansion of the relative dielectric constant at $p_0$ (atmospheric pressure) and $T_0 = 273.15$ K [25]. $u_{ij}$ is the $j$-th root of the $i$-th order Bessel function of the first kind, and $a(\omega) = a_{AP} + 4\pi^2 c^2 q/(\omega^2 t)$ is the frequency-dependent effective core radius, where $a_{AP}$ is the area-preserving core radius, $t$ is the core-wall thickness and $q = 0.065$ is an empirical parameter derived from fitting Eq. (1) to the results of finite element modeling of an evacuated, ideal kagomé fibre [24]. The dephasing parameter for $LP_{01}$ to $LP_{ij}$ conversion takes the form:

$$\vartheta_{ij} = \beta_{AS}^{ij} + \beta_S^{ij} - 2\beta_P^{01} \qquad (2)$$

and perfect phase-matching ($\vartheta_{ij} = 0$) can be arranged by appropriately adjusting the pressure of the filling gas (the weak anomalous dispersion of the hollow core modes is counterbalanced by the normal dispersion of the gas [26]).

In this work we study the dominant vibrational mode of hydrogen, which has a Raman frequency shift of $\Omega_R = 125 \times 10^{12}$ Hz. At $T = 298$ K, the vibrational Raman gain coefficient $g_P$ [m/W] at pump frequency $\omega_P$ [Hz] and pressure $p$ [bar] can be written [27]:

$$g_P(\omega_P, p) = \frac{9.37 \times 10^{12} (57.2 p/\Delta\nu)(\omega_P - \Omega_R)}{c(7.19 \times 10^{13} - \omega_P^2/c^2)^2}, \qquad (3)$$

where $\Delta\nu = (280/p) + 56.98 p$ is the Raman linewidth in MHz. For pressures above ~10 bar, $g_P$ saturates to maximum values of 2.94 cm/GW and 9.46 cm/GW at 532 nm and 266 nm pump wavelengths respectively.

The effective Raman gain coefficient in the fiber, $\gamma_{eff}$ [m/W], is defined as $\gamma_{eff} = \rho_{ij} S_{ij} g_P$, where $S_{ij}$ is the spatial overlap integral between the $LP_{01}$-like pump mode and the $LP_{ij}$-like Stokes mode and $\rho_{ij}$ is the gain reduction factor [23]:

$$\rho_{ij} = \left| \text{Re}\sqrt{\left(\frac{q-1}{2}\right)^2 - \frac{\vartheta_{ij}}{g_P I_P}\left(\frac{\vartheta_{ij}}{g_P I_P} + i(q+1)\right)} - \left(\frac{q-1}{2}\right) \right| \qquad (4)$$

where $q = g_{AS}\omega_{AS}/g_P\omega_P \sim 1.65$ at a pump wavelength of 532 nm.

Eq. (4) shows that, for conversion to the $LP_{01}$ mode, the gain reduction factor depends on the ratio $|\vartheta_{01}/(g_P I_P)| = \pi(L_G/L_D)$, where $L_G$ is the exponential gain length and $L_D = \pi/\vartheta_{01}$ the dephasing length. The gain suppression factor equals 0.83 at $\vartheta_{01}/g_P I_P = 2$, rapidly dropping for smaller values of $\vartheta_{01}/g_P I_P$. The long low-loss single-mode interaction lengths in ARR-PCF make exploration of this regime straightforward, in contrast to previous work with bulk gas cells [19].

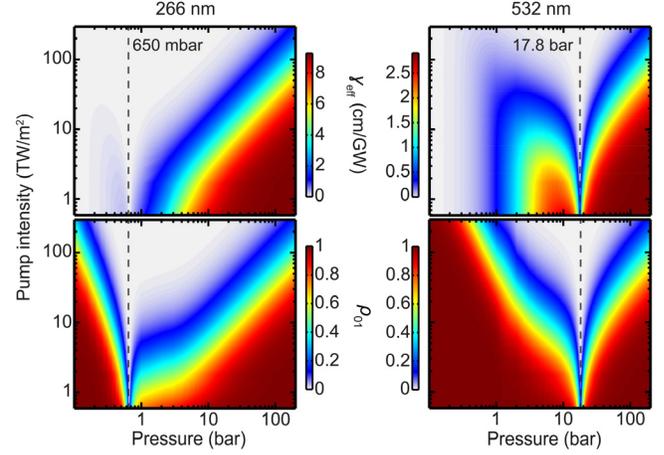

Fig. 1. Effective Raman gain coefficient $\gamma_{eff}$ (upper) and gain reduction factor $\rho_{ij}$ (lower) for the $LP_{01}$ mode, plotted as a function of pump intensity and gas pressure for pump wavelengths 266 nm (left) and 532 nm (right). The dashed lines indicate the position of the phase-matching pressure for a $H_2$-filled kagomé-type ARR-PCF with 21.3 μm core diameter. Due to a higher Raman gain at the shorter wavelength, the pressure range where $\rho_{01} < 1$ is wider for the UV pump.

To illustrate the effect, $\gamma_{eff}$ and $\rho_{01}$ are plotted as functions of gas pressure and pump intensity in Fig. 1 for a hydrogen-filled kagomé-PCF with a flat-to-flat core diameter of 21.3 μm. For pump wavelengths of 266 nm and 532 nm, the pressures $p_{01}$ for perfect gain suppression ($\vartheta_{01} = 0$) are ~650 mbar and ~17.8 bar, respectively. At fixed pressure, $\rho_{01}$ drops with increasing pump intensity as expected from Eq. (4), whereas for a fixed pump intensity the pressure range over which gain reduction effects are important broadens. This effect can be quantified using the normalized parameter $\Delta p = |(p_{1/2} - p_{01})/p_{01}|$ where $p_{1/2}$ is the pressure ($> p_{01}$) at which $\rho_{01} = 0.5$. In Fig. 2(a) $\Delta p$ is plotted against pump intensity, showing how dramatic is the effect of higher Raman gain in the ultraviolet: the pressure range for gain suppression is more than 10 times larger.

Figure 2(b) shows the dispersion curves of the $LP_{01}$ mode for different gas pressures. Sufficiently away from $p_{01}$, for example at 10 bar for a 266 nm pump and 30 bar for a 532 nm pump, the ZDP is located far from the central pump wavelength so that $\vartheta_{01}$ is far from zero. Although under these conditions one might expect coherent gain suppression to be very weak [28], this turns out not to be the case at high pump intensities, because the overall gain per unit length, $G_{ij}$, is proportional to the product of $\rho_{01}$ and $I_P$. Indeed, $\rho_{01}$ can dominate over $I_P$ at high enough pump intensity, resulting in strong gain suppression. This is illustrated in the left-hand panel of Fig. 3(a), where $G_{01}$ is plotted against pressure and pump intensity using the same parameters as in Fig. 1 for 532 nm pump wavelength. As $I_P$ increases, higher gas pressure is required to maintain high gain.

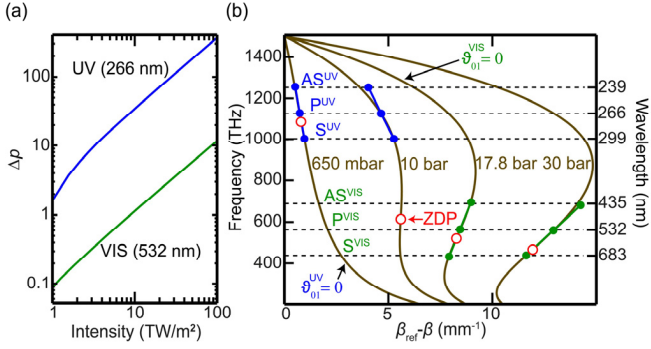

Fig. 2. (a) $\Delta p$ as a function of pump intensity. (b) Dispersion diagram of the $LP_{01}$ mode for a kagomé-style ARR-PCF with 21.3 μm core diameter at different hydrogen-filling pressures. Horizontal dashed lines indicate the positions of pump (P), Stokes (S), and anti-Stokes (AS) bands. The solid-blue and green lines indicate the coherence waves present in the system. The red circles indicate the position of the ZDP for each pressure. In order to highlight the small changes in the dispersion, the frequency-dependent parameter $\beta_{ref} = (\omega/\omega_m)\beta(\omega_m)$, $\omega_m$ being the maximum plotted frequency, is subtracted from $\beta$.

For comparison, the gain in the absence of gain suppression, $G_{01}/\rho_{01}$, is also plotted (right-hand panel). Figure 3(b) plots $G_{01}$ and $G_{01}/\rho_{01}$ against increasing $I_P$ at a fixed pressure of 40 bar, for which $\vartheta_{01} \sim 300$ m$^{-1}$ ($p_{01} = 17.8$ bar). $G_{01}/\rho_{01}$ increases linearly with pump intensity as expected. $G_{01}$, however, initially follows $G_{01}/\rho_{01}$ but then begins to saturate as $\rho_{01}$ starts to play a role. As discussed above, $G_{01}$ actually drops at even higher values of $I_P$, as a result of the dominance of gain suppression. Interestingly, above a certain pressure-dependent intensity threshold the intermodal $LP_{01}$-$LP_{11}$ gain $G_{11}$ (full green line), which is unaffected by gain suppression at the pressures achievable in the experiments ($\vartheta_{11} = 0$ at 490 bar), overtakes the intramodal $LP_{01}$-$LP_{01}$ gain, causing the Stokes signal to be emitted predominantly in the $LP_{11}$ mode (see next section).

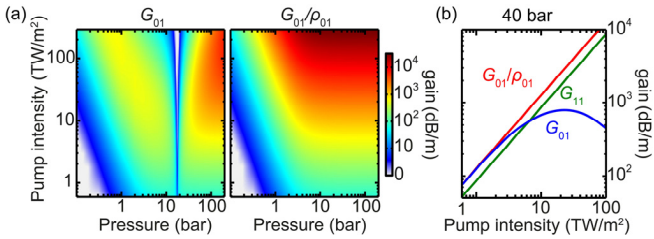

Fig. 3. (a) Left-hand panel: overall Stokes gain per unit length $G_{01}$ (theory) plotted on a logarithmic scale against pump intensity and gas pressure. Right-hand panel: the Stokes gain with the gain suppression factor removed, $G_{01}/\rho_{01}$. The pump wavelength is 532 nm and the other parameters are identical to those used in Fig. 1. (b) Intramodal gain $G_{01}$ (full blue curve) and intermodal gain $G_{11}$ (full green curve) plotted against pump intensity at a pressure of 40 bar. $G_{11}$ shows no sign of gain suppression at this pressure ($p_{11} = 490$ bar). The full red curve plots $G_{01}/\rho_{01}$ for comparison.

## III. EXPERIMENT

The experimental set-up for observing intensity broadening of the gain suppression region is sketched in Fig. 4. A length of kagomé-PCF was attached to two pressure cells, evacuated and then filled with hydrogen to the required pressure. Pump pulses at 532 nm (1 ns) and 266 nm (3 ns) were launched into the core, and the resulting Stokes signal separated from the pump light using a prism. The near-field mode profile of the Stokes light was imaged using a CCD camera, and the Stokes power measured with a power meter.

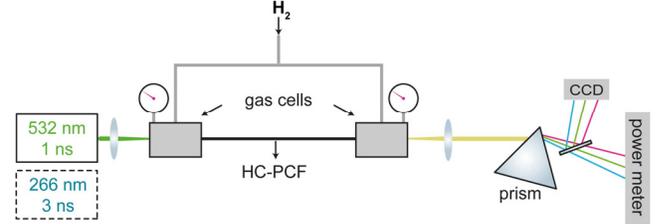

Fig. 4. Schematic of the experimental set-up. Two different pump lasers, at 532 and 266 nm, were used for the experiments. Linearly-polarized pump light was used to prevent generation of rotational sidebands. Stokes light was spatially separated from the pump using a prism. Two different kagomé-PCFs were used in the experiment: for the visible pump, a 37-cm-long fiber with a core diameter of ~21.3 μm, and for the UV pump a 10-cm-long fiber with a core diameter of ~23 μm.

### A. Pumping at 532 nm

Pulses at 532 nm were launched into a 37-cm-long $H_2$-filled kagomé-PCF with 21.3 μm core diameter. The pressure was gradually increased from 6 to 32.5 bar and the Stokes power at 683 nm was measured, along with its near-field mode profile, for launched pulse energies of 1.80, 2.25 and 3.60 μJ. In the experiments it was difficult to avoid launching a small fraction of the pump beam into the $LP_{11}$ mode. This resulted in a weak $LP_{11}$ Stokes signal through scattering off the $LP_{01}$-$LP_{01}$ coherence wave. Although this conversion process is phase-mismatched, it produces an $LP_{11}$ Stokes signal strong enough to seed $LP_{01}$-$LP_{11}$ conversion, resulting in a strong $LP_{11}$ Stokes signal at the fiber output.

Figure 5(a) plots the Stokes pulse energies measured at different pressures and Fig. 5(c) shows the corresponding near-field images. The red points correspond to conditions where inspection of the near-field images revealed noticeable $LP_{11}$ mode content in the Stokes signal, i.e., when there was significant $LP_{01}$-mode gain suppression. This region widens as the pump energy increases. Comparing points (2) and (6) in Fig. 5(c), we observe that the $LP_{11}$ mode content also increases with intensity, providing evidence of decreasing overall intramodal gain $G_{01}$. This has important practical implications for the design of Raman HC-PCF-based lasers and amplifiers, where a high output power, as well as a pure $LP_{01}$ mode profile, are desirable.

This seemingly negative aspect of gain suppression effect can be turned to advantage if the Stokes signal is required to emerge in a particular higher order mode for applications in, e.g., optical tweezers or particle trapping [29]. By choosing an appropriate pressure and pump intensity, $LP_{01}$-$LP_{01}$ Stokes generation can be strongly suppressed while $LP_{01}$-$LP_{11}$ conversion is strongly enhanced (provided competing processes, such as generation of a second Stokes band or other intermodal transitions, do not intervene).

To emphasize the universality of coherent Raman gain suppression, the theoretical value of $\rho_{01}$ is plotted in Fig. 5(b), overlaid with experimental data-points. The dashed-brown lines enclose the region where the effective Raman gain of the $LP_{11}$ mode $g_P\rho_{11}S_{11} > g_P\rho_{01}S_{01}$ ($\rho_{11} = 1$ for the pressure range investigated (see above), $S_{01} = 1$ and $S_{11} = 0.68$) [23]. There is excellent agreement between experiment and the analytical predictions of Eq. (4).

### B. Pumping at 266 nm

Coherent gain suppression is more pronounced at pump wavelengths in the UV, when the Raman gain $g_P$ is much higher (see above). To investigate this, narrowband 266 nm pulses (4$^{th}$ harmonic of the 1064 nm Q-switched laser) with ~3 ns duration were launched into a 10-cm-long kagomé-PCF with ~23 µm core diameter. Under these conditions $p_{01} = $ ~550 mbar [16], so that naïvely one might expect that there should be no trace of gain suppression effects at high pressure. This is however far from the truth, as we now discuss.

Fig. 6 shows near-field optical images of the UV Stokes signal (emitted at 299 nm) at different pressures when pump pulses with ~10 µJ energy were launched into the fiber. It is evident that, below ~14 bar, the Stokes signal is emitted in a mixture of HOMs, indicating that the $LP_{01}$-$LP_{01}$ gain is significantly lower than the intermodal gain below this pressure.

A simple estimate using Eq. 4 gives a transition pressure of ~16.7 bar, close to the experimental value. Thus, below this threshold the UV intramodal SRS will be very inefficient—a counterintuitive situation if coherent gain suppression is not taken into account.

In the UV experiment at lower pressures the near-field distribution of the Stokes signal suggests that it is in a superposition of $LP_{02}$ and $LP_{21}$ modes (bottom row in Fig. 6), i.e., that intermodal $LP_{01}$-$LP_{02}$ and $LP_{01}$-$LP_{21}$ pump-Stokes conversion is taking place simultaneously, perhaps aided by the launching of small fractions of HOM pump light, as already mentioned above.

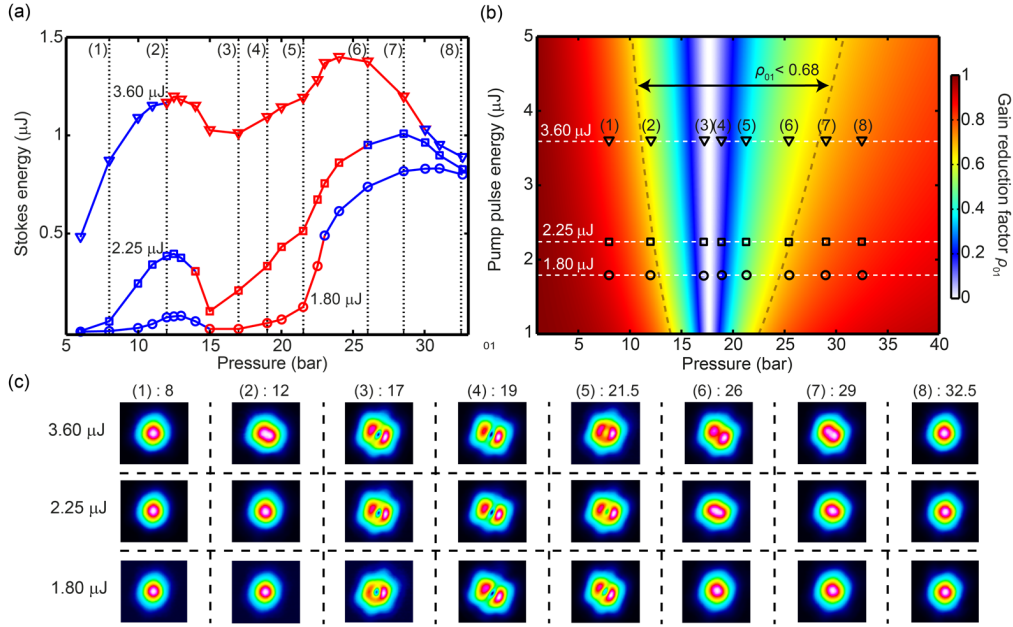

Fig. 5. (a) Stokes power measured for 1 ns pump pulses, wavelength 532 nm, with energies of 1.80, 2.25, 3.60 µJ. The gas pressure is varied from 6 to 32.5 bar. The dashed lines indicate the pressures at which the near-field Stokes profiles were imaged (shown in (c)). The pressures at which the Stokes is emitted predominantly in the $LP_{11}$ mode are plotted in red. (b) Gain reduction factor $\rho_{01}$ plotted as a function of energy and pressure. The numerals correspond to the experimental points marked in (a) and (c). The effective Raman gain of the $LP_{01}$ mode is lower than that of the $LP_{11}$ mode within the region bounded by the brown dashed lines. (c) Near-field profiles of the Stokes beam at each pressure and pulse energy, recorded with a CCD camera. The pressures in each case are shown at the top of the images.

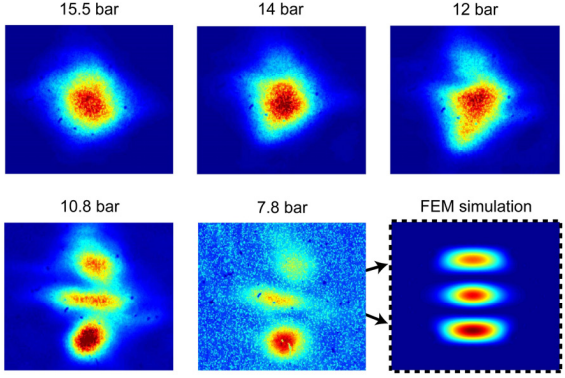

Fig. 6. Near-field images of the UV Stokes beam at different pressures. The modal pattern converges towards a $LP_{01}$-like mode at sufficiently high pressure. The lower right-hand plot shows the finite element modeling of the near-field mode profile at the Stokes frequency for a mixture of $LP_{02}$ and $LP_{21}$-like modes; it closely resembles the experimental measurements at 10.8 and 7.8 bar.

## IV. MODELING AND DISCUSSION

To accurately describe the Raman interaction under conditions of coherent gain suppression, we have made use of the coupled Maxwell-Bloch equations described previously [23,30], for the case when the majority of the molecules are in the ground state. Using these equations, coherent Raman gain suppression can be classically explained as the mutual cancellation of the optical fringe patterns created by pump-Stokes and pump-anti-Stokes interference: these field patterns drive the creation of Raman coherence waves. When only $LP_{01}$ pump, Stokes and anti-Stokes fields are considered in the steady-state regime (when the pulse duration is much longer than $T_2$) [31], the driving fringes produces a coherence wave:

$$Q = -i\frac{T_2}{4}(\kappa_{1,1}|E_1|e^{i\phi_1}|E_0|e^{-i\phi_0}q_1q_0^* + \kappa_{1,0}|E_0|e^{i\phi_0}|E_{-1}|e^{-i\phi_{-1}}q_0q_{-1}^*). \quad (5)$$

where $E_l$ is the slowly varying field amplitude of the $l$-th sideband, $\phi_l$ its phase, $q_l = \exp[-\beta(\omega_l)z]$ and the coupling constants are defined by:

$$\kappa_{1,l} = \frac{1}{Z_0}\sqrt{\frac{2g_l}{NT_2\hbar\omega_{l-1}}}, \quad \kappa_{2,l} = \sqrt{\frac{g_lN\hbar\omega_{l-1}}{2T_2}}. \quad (6)$$

where $N$ is the molecular number density, $\hbar$ the reduced Planck's constant, $Z_0$ the impedance of free space and $T_2$ is the dephasing time of the Raman polarization [31]. The values of $\kappa_{1,l}$ and $\kappa_{2,l}$ can be obtained from the experimental gain values $g_l$ [27].

For complete gain suppression, the two contributions on the right-hand side of Eq. (5) must cancel out. This occurs when the amplitudes and phases of the interacting fields at pressure $p_{01}$ (when $\vartheta_{01} = 0$) satisfy:

$$\begin{cases}\Phi = (2n+1)\pi \\ \kappa_{1,1}|E_1(z,\tau)| = \kappa_{1,0}|E_{-1}(z,\tau)|,\end{cases} \quad (7)$$

where $\Phi = \phi_{-1} + \phi_1 - 2\phi_0 + \vartheta_{01}z$ is the global phase [32], $n$ is an integer and $\kappa_{1,1}/\kappa_{1,0} \sim 1$ in the visible domain. Gain suppression ($Q \sim 0$) will occur if the Stokes/anti-Stokes fields have similar amplitudes and the global phase is an odd multiple of $\pi$. If these conditions are already met at the fiber input, stimulated Raman scattering is inhibited and only incoherent spontaneous scattering will be present. In contrast, if the input global phase is externally set to a different value, the Raman coherence will be initially non-zero, enabling the transfer of energy between the fields until the effect of phase-locking stabilizes the global phase to a multiple of π and the gain stops [32]. In practice, this phase stabilization effect occurs over millimeter length-scales, much shorter than the fiber length, resulting in negligible conversion to noise-seeded Stokes and anti-Stokes bands [33].

To model the dynamics of the real system, we solved the Maxwell-Bloch equations including both $LP_{01}$ and $LP_{11}$ fiber modes. The Stokes mode energies for a 532 nm pump are plotted against gas pressure in Fig. 7, showing good agreement with the experimental results. The only free parameters, adjusted for the best agreement with the experiments, are the initial Stokes sideband amplitudes (200 V/m in Fig. 7(a) and 100 V/m in Fig. 7(b)) and the loss of the $LP_{11}$ mode, which we take to be five times that of the $LP_{01}$ mode (a reasonable assumption based on results obtained for similar fibers [13,34]).

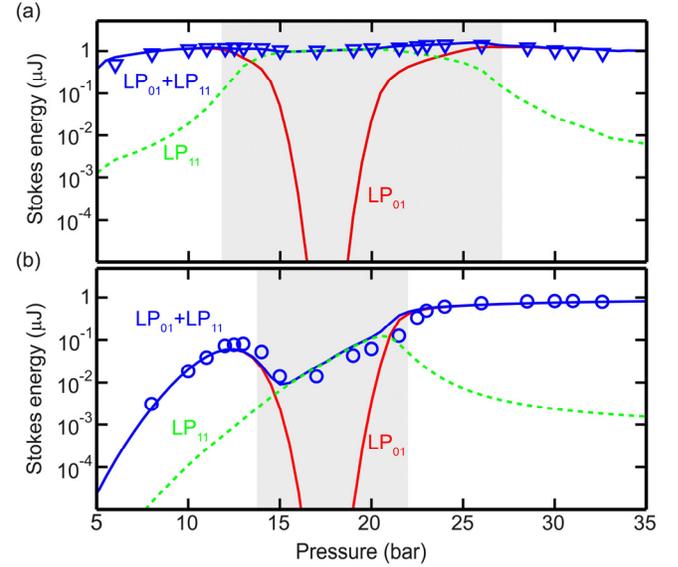

Fig. 7. Simulated Stokes output energy for (a) 3.60 µJ and (b) 1.80 µJ pump energy. The full red line represents the Stokes signal generated in the $LP_{01}$ mode and the dashed-green line the signal in the $LP_{11}$ mode. The full-blue line is the sum of the two modal contributions. The experimental data points (same as in Fig. 5) are also shown with blue symbols.

Given the good agreement, we can extract information from the simulations about the precise modal content of the different signals (something that was experimentally inaccessible). In Fig. 7 we observe that the region where the $LP_{11}:LP_{01}$ Stokes energy ratio exceeds 1:10 (indicated by a gray-shaded area) is clearly wider for higher input pulse energies. In fact, the regions where the $LP_{11}$ Stokes contribution is dominant agree well with our estimates (the red data-points in Fig. 5(a)) based on examination of the near-field profiles in Fig. 5(c).

In Fig. 8 we plot the simulated spatio-temporal evolution of the inter/intramodal coherence at a pressure of 35 bar (top panels) and 17.5 bar (bottom panels), for 1.80 μJ pump energy. It is evident that the intramodal coherence is suppressed by more than five orders of magnitude at 17.5 bar, i.e., when operating close to $p_{01}$ (bottom panel in Fig. 8(a)). As a consequence, intermodal SRS to the $LP_{11}$ mode is dramatically enhanced and indeed dominates the Stokes signal. Interestingly, and in contrast to the 35 bar case, the intermodal coherence continuously increases along fiber due to a lower effective gain. At 35 bar there is evidence of the onset of a second Stokes band in the form of a longitudinal modulation of the intramodal coherence (top panel of Fig. 8(a)). This modulation is caused by interference between the intramodal coherence waves generated by the pump/first-Stokes and first/second-Stokes signals [26]. The beat-length of this fringe pattern is

$$L_B = \frac{2\pi}{\left|2\beta_{S1}^{01} - \beta_P^{01} - \beta_{S2}^{01}\right|} = 23.2 \text{ mm} . \quad (8)$$

Although not shown here, the model described in this section may also be applied for UV pumping [16], where similar results are expected.

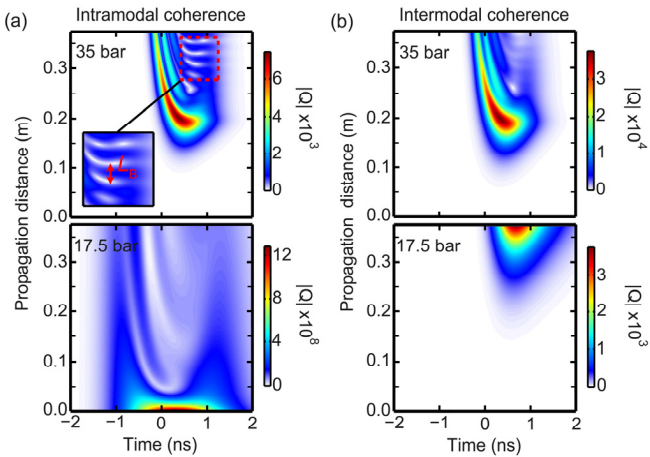

Fig. 8. Spatio-temporal evolution of the steady-state (a) intramodal ($LP_{01}$-$LP_{01}$) and (b) intermodal ($LP_{01}$-$LP_{11}$) coherence at a pressure of 35 bar (top panels) and 17.5 bar (bottom panels). The pump pulse energy is 1.80 μJ. The inset shows a detail of the interference fringes observed in the intramodal coherence at high pressure.

## V. SUMMARY AND CONCLUSIONS

In summary, intramodal Raman gain in gas-filled HC-PCFs is always significantly suppressed at high pump energies, even when operating far from the zero-dispersion point. Coherent gain suppression is stronger at shorter pump wavelengths, when the material Raman gain is larger. This is a key effect that should be taken into account in the design of Raman frequency shifters, amplifiers and frequency comb generators based on gas-filled broadband-guiding HC-PCFs, or in any application requiring a high conversion efficiency to Raman sidebands emitted in a pure $LP_{01}$ mode. This is because HC-PCFs inherently support HOMs with losses that, although high, can easily be overcome by the very large intermodal Raman gain made possible by intramodal ($LP_{01}$-$LP_{01}$) gain suppression. The result is high conversion efficiencies (~30%, limited by generation of a strong second Stokes signal) to the $LP_{11}$ mode. Coherent gain suppression offers the unique possibility of generating a Stokes beam in a complex higher order mode, with potential applications in, e.g., supercontinuum generation [35], optical tweezers [36] and particle trapping [29].